# In situ transformation and cleaning of tin-drop contamination on mirrors for extreme ultraviolet light

Norbert Böwering[a)]

BökoTech, Ringstr. 21, 33619 Bielefeld, Germany, and
Molecular and Surface Physics, Bielefeld University, 33615 Bielefeld, Germany

Christian Meier

Molecular and Surface Physics, Bielefeld University, 33615 Bielefeld, Germany

[a)] Electronic mail: boewering@physik.uni-bielefeld.de

## Abstract

Tin-drop contamination was cleaned from multilayer-coated mirrors by induction of phase transformation. The β→α phase transition of tin was induced to initiate material embrittlement and enable facile removal of thick tin deposits. The necessary steps were performed under high-vacuum conditions for an in-situ demonstration of the removal of severe tin contamination from optics used for reflection of extreme ultraviolet light. Molten tin of high purity was dripped onto mirror samples, inoculated with small seed particles of gray tin and then cooled to temperatures in the range of -25 °C to -40 °C. As recorded by photographic imaging, the drops were converted in an evacuated chamber to gray tin by induction of tin pest leading to their disintegration within a few hours. They could then be easily cleaned or fell off from the surface without causing any damage of the multilayer coating. Cleaning of tin contamination from the mirrors with almost complete structural transformation of the tin drops and subsequent removal by puffs of dry gas could be achieved within a day. The fraction of area coverage of untransformed tin remaining on the samples after cleaning was evaluated from the images and generally found to be well below 1%. After tin dripping, phase transition and cleaning, analysis of the reflectance of a Mo/Si-coated mirror with measurements at wavelengths of 13.6 nm and 13.5 nm showed a reduction by only 0.5%, with an upper limit of 1%.

# I. INTRODUCTION

In recent years, impressive progress has been made in the development and industrialization of high-performance scanners for next-generation lithography at extreme ultraviolet (EUV) wavelengths to be used in semiconductor manufacturing.[1] The laser-produced plasma light sources for EUV generation operated with tin droplet targets[2] have now reached output powers above 200 W that will enable high-volume manufacturing at chipmakers at throughputs exceeding the minimum requirement of ~125 wafers per hour.[3] To maintain high productivity levels of the expensive EUV scanner machine, not only the power of the light source but also its availability and uptime are equally of crucial importance. A considerable part of the availability budget of the source is taken up by the light collection mirror that is located near the bottom of the source chamber in very close vicinity to the hot tin plasma. It collects the generated EUV light at around 13.5 nm but can also accumulate considerable amounts of tin debris leading to deterioration of its reflectance over time. This large optics reflects at near normal incidence angles by means of a multilayer (ML) coating of ~50 alternating Mo and Si layers with a typical bilayer spacing of about 7 nm providing Bragg reflection of the incident EUV light from the tin plasma towards the illumination optics of the scanner.[4,5] Although collector lifetimes exceeding 100 Billion pulses have been demonstrated at lower EUV power levels in source development as early as 2014[6], further progress is urgently required to reach lifetimes greater than a few months during 24/7 high-power operation at repetition frequencies of ~50 kHz or more for chip production. Source debris mitigation techniques based on flows of hydrogen buffer gas that are employed for lifetime extension are only partially effective at the low pressure environment needed to transmit the generated EUV radiation.[3,6] Keeping the gas absorption at acceptable levels limits the pressure in the source chamber. The stopping of energetic ions and of tin microparticles by collisions with hydrogen is then not complete. This can lead to buildup of tin contamination on the collector mirror. Furthermore, tin deposited on walls inside of the source chamber can fall or drip onto the collector surface.[3] Therefore, cleaning of accumulated Sn contamination from the mirror surface is periodically required in order to maintain the collector reflectivity at acceptable levels for long times during light source operation.

The collector mirror is an expensive consumable optics component. Collector modules in industrial sources are therefore designed for exchange at end-of-life enabling later use after cleaning and refurbishment processes. In the past, several different methods have been examined and tested for cleaning contaminating Sn deposits from the mirror surface. Mechanical techniques for tin removal are not considered to be viable cleaning options because of the high risk of induced coating damage. A wet-etch technique using as a cleaning agent an acid solution like HCl (hydrochloric acid) has been described.[7] However, this method usually requires complete disassembly of the collector module and its reassembly followed by qualification after mirror cleaning, resulting in turn-around times of several weeks. More extensive refurbishment by chemical removal of the top multilayers or entire coating from the EUV mirrors followed by re-coating is possible, but even more expensive and more time-consuming.[8,9] As a considerably faster possibility for optics cleaning, the etching of tin by hydrogen radicals was examined in detail.[10-13] It leads to the formation of volatile $SnH_4$ (tin tetra-hydride) molecules that may be pumped away. Corresponding schemes of plasma dry etching were successfully tested for collector mirrors with in-situ cleaning demonstrations carried out inside of the vacuum system, potentially avoiding a lengthy collector swap with related system downtimes.[6,13] However, due to limited etch rates, reasonable cleaning times can only be achieved for the removal of Sn layers with a thickness of up to several tens of micrometers. It seems that thicker and more massive tin depositions are not easily removable from EUV optics in this way. Dry cleaning methods using intense gaseous or frozen-particle flows like $CO_2$ snow pellets, as described for EUV light source or mask cleaning, may also be applied.[14-16] However, such gas jetting techniques are most suitable for the removal of rough and powdery layers and less effective for thick and smooth Sn deposits adhering well to the coated mirror surface. Furthermore, this is a cleaning scheme that can only be applied to the collector when the source chamber is fully vented. In this case, the collector module might as well be swapped for carrying out an ex-situ cleaning process.

In this paper, we describe and examine an alternative cleaning technique for tin-contaminated EUV optics based on the initiation of a phase transformation of the deposited tin prior to its removal by puffs of dry or inert gas. The basic concept for cleaning tin-drop contaminated optics in this way, and the related material science aspects of this method were presented very recently in a communication by one of us.[17] When tin in its normal metallic phase, β-Sn, is cooled to temperatures in a range of about -20 °C to -50 °C, it can undergo a phase transformation to gray tin, a rather brittle semimetal.[18] This phase, α-Sn, is the stable phase of tin with slightly lower entropy in comparison to β-Sn at temperatures below 13 °C.[19] The transition, also called tin pest, occurs by a process of nucleation and growth[18,20]; generally, it needs to be induced by a seeding initiation mechanism on the surface[21]. As illustrated in Fig. 1, the allotropic transformation is accompanied by a structural change of the material from a body-centered tetragonal to a face-centered cubic (diamond-like) lattice leading to a strong volume expansion of about 26%. The extreme brittleness of converted tin deposits and its breakup by cracking after transformation to the α-Sn phase is the attractive feature for the purpose of optics cleaning.

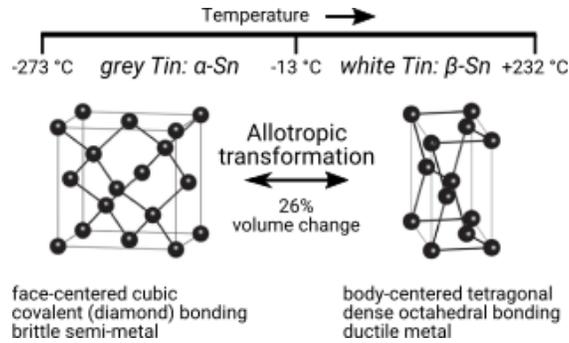

Fig. 1. Illustration of temperature-dependent allotropic transformation between α-phase gray tin and β-phase white tin. The corresponding crystal structures are shown, as well as the geometric classification, the bonding and the metallic nature for both phases.

In the previous work[17] the transformation of tin drops was studied with its dependence on material purity as well as its use for ex-situ cleaning of ML-coated EUV mirror samples from Sn drops in air. However, cleaning should ideally be done in situ *inside* of the source chamber to avoid long system downtimes originating from venting, pumping and module swap cycles. Reliable and fast in-situ removal of thick tin deposits from the collector would greatly increase the source availability for the EUV scanner. It would also represent a significant cost reduction factor, if collector swaps could be avoided and the number of spare collector modules per system could be decreased in this way. Therefore, the main focus of the present study is on testing the tin cleaning by induction of tin pest inside of an ultra-high vacuum (UHV) chamber. In contrast to previous studies where the etching of deposited Sn layers was examined for film thicknesses up to 200 nm[10,11,13], as a worst-case contamination scenario we study here massive Sn drops with a thickness of many micrometers, or even up to millimeters, by in-situ dripping of molten tin onto ML-coated optic samples. The purpose is to examine how β-Sn deposits can be transformed in situ under vacuum conditions to the α-Sn phase for facile subsequent and damage-free removal from the surface of EUV mirrors. The phase transition on several ML-coated mirrors with tin drop contaminations was studied during their thermal contact with a cooled sample holder. Inoculation of the Sn drops by small seed particles of α-Sn was required for fast conversion. The results for in-situ β→α Sn phase transformations on two representative samples are presented and discussed in detail in the results section of this paper. In addition, to analyze potential influences on mirror reflectivity, EUV reflectance data obtained from a sample before and after tin cleaning are reported.

## II. EXPERIMENTAL SETUP, VACUUM EQUIPMENT AND METHODOLOGY

An UHV chamber with flanges of up to 8-inch size that had previously been used for tests of ML-coating depositions by electron-beam evaporation was adapted for use in tin transformation experiments. The main parts of the experimental setup are shown schematically in Fig. 2. The vacuum chamber consisted of a vertically mounted, 8-inch diameter cylinder with preparation and diagnostic capability in its upper section and sample holding and cooling part in its lower section. In the upper portion, a heatable tin drip tray could be inserted via a 2-inch vacuum translation stage. An ionization gauge and a quadrupole mass analyzer were attached to the chamber for vacuum diagnostics. Sample illumination was provided through the top and in addition (optionally) also through vacuum windows on the side of the chamber. Optical detection was the main analysis method. It enables the monitoring of the behavior of tin contamination during phase transition in situ by illumination, imaging and recording through a vacuum window. A digital camera (Canon EOS 350D, 8 megapixel) with Zoom objective mounted on axis at the top on a tripod was used for recording images of the samples at predefined intervals during transformation experiments.

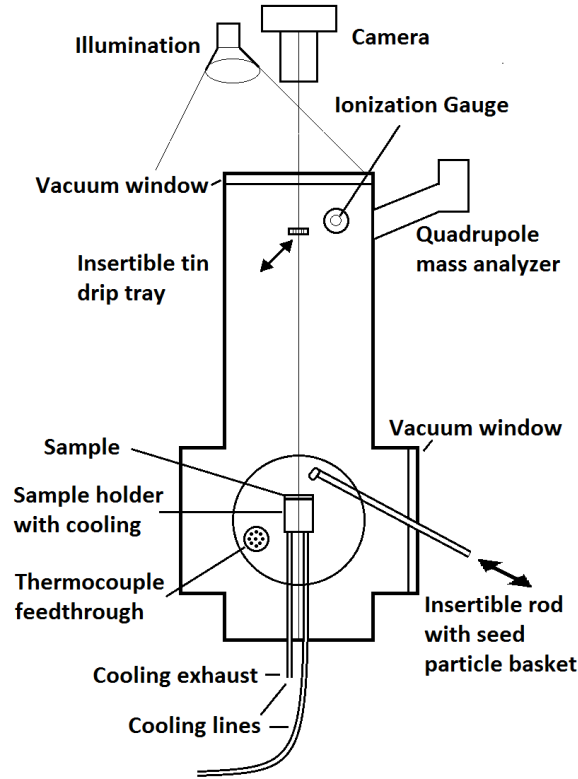

Fig. 2. Schematic view of the experimental setup. The vacuum chamber contained the equipment for tin dripping, sample inoculation, substrate cooling and temperature measurements. Illumination and photographic recording was provided via vacuum windows.

The ML-coated EUV mirror samples were mounted in horizontal orientation on a fixed holder made from stainless steel and copper blocks with attached cooling lines inserted at one of the four lower 6-inch vacuum flanges. The copper table on which samples could be mounted had an area of 50 mm x 50 mm. A multiple-connection vacuum feedthrough enabled temperature monitoring by several sets of thermocouple wires (chromel-alumel type K) that could be attached at various positions on the sample or sample holder. A magnetically-coupled linear translation rod with rotation possibility was mounted via adjustable bellows for manipulation for insertion at a distance of a few centimeters above the samples. By means of small baskets it provided the capability for in-situ inoculation with tin seed particles under vacuum conditions by dropping α-Sn powder onto the sample surfaces.

A liquid-nitrogen container with pressure regulation and relief valves was attached to the inlet of the cooling line. It enabled a simple means of sample cooling by a regulated flow of cold nitrogen vapor from evaporated liquid nitrogen leading to pressure-built up inside of the reservoir. Temperatures measured by thermocouple probes at the sample holder could be kept stable to within ±5 °C by this technique. Figure 3 shows a corresponding photo of the UHV chamber with its attached equipment. The chamber was pumped by a 330 l/s plasma turbo-molecular pump (Balzers TPH 330S, not visible in the photo) operated with fomblin lubricant and a rotary two-stage fore-pump equipped with a trap for oil-vapor suppression. For protection of the blades of the turbo-pump from metal particles a large baffle trap was positioned at the backside pumping port between the vacuum chamber and pump section. The base pressure of the UHV chamber was about $10^{-5}$ Pa when its walls were not baked. A mix of copper and viton gaskets was used for the sealing of the conflat vacuum flanges.

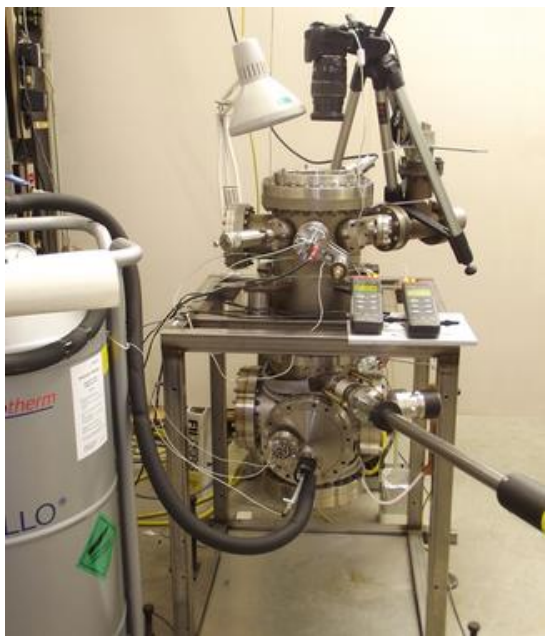

Fig. 3. (Color online) Photograph of vacuum chamber used during tin transformation experiments. A liquid-nitrogen container was connected via insulated cooling line for sample cooling. Photographic imaging and temperature recording equipment is also visible.

A photo of the tin drip tray is shown in Fig. 4a. Before vacuum pump-down, several pieces of high-purity tin (previously untransformed β-Sn of 99.999% purity, mass of each piece: ~0.3 g) were placed on its two slots. Aluminum was used as tray material, since molten tin does not adhere to it and no alloys are formed at the melting temperature of tin. For the melting and dripping of tin, the tray was slowly heated up from room temperature to temperatures in the range of about 255 – 265 °C by resistive electrical heating while the tray temperature was monitored by a thermocouple probe. After melting, the tin pieces fell down as liquid drops if the adhesion to the tray was not too strong. If required, small vibrations to induce the dripping of tin could be created by touching the tray support with a bar mounted on a rotation feedthrough. After completion of tin dripping the tray was retracted in order to remove it from the view of the camera. The distance from the trip tray to the samples below was 41 cm, leading to a tin droplet speed of 2.84 m/s at impact. After full melting, flat tin drops were typically produced on the mirror samples with diameters in the range of 6 mm to 8 mm, often exhibiting a "finger" pattern at the edge. For single drops of fully melted tin the thickness at the center of the drop was measured to be typically about 0.30 – 0.35 mm. The relative surface area coverage by tin contaminations of the samples after tin dripping and after cleaning was determined from photos by means of software for image-analysis (ImageJ).

Small pieces of Si(100) wafers (less than 1 inch size) with the surface polished to a roughness of <0.2 nm root-mean-square were used as EUV mirror substrates. They had previously been ML-coated by electron-beam evaporation.[22] The coatings consisted of alternating Mo and Si layers forming multilayers (50 bilayers) optimized for a peak reflectance $R_{max}$ above 0.65 at a wavelength close to 13.5 nm at near normal incidence. The sample holder could accommodate either two square mirror samples with about 20 mm x 20 mm exposed area or up to six rectangular mirror samples of ~10 mm x 20 mm, held in place by a corresponding mounting plate made from aluminum. The EUV mirror substrates were cooled on the bottom side by thermal contact with the copper cooling table via contact pressure established by the Al holder. The EUV reflectance of a ML-coated sample was measured at 5° incidence angle at the reflectometer of the Physikalisch-Technische Bundesanstalt (PTB) in Berlin using synchrotron radiation.[23]

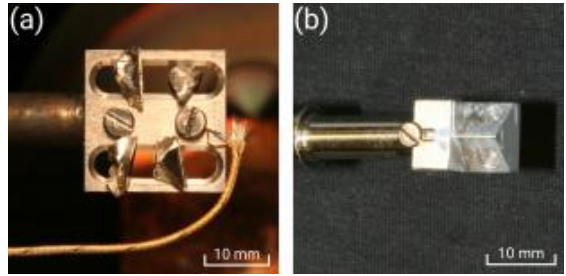

Fig. 4. (Color online) (a) Heatable, slotted drip tray with tin pieces for melting and dripping. A thermocouple sensor was attached to the tray to monitor the temperature. (b) Tip of manipulator rod with two small baskets for sample infection.

The tip of the manipulator rod for in-situ inoculation is shown in the photo of Fig. 4b. It consisted of an aluminum piece with two bores of 3 mm diameter creating small baskets that could be loaded with α-Sn powder for sample infection. By adjusting the bellows of the manipulator and rotating the baskets either clock- or counterclockwise for emptying, the tin contamination on the EUV mirror samples could be inoculated in situ from above with α-Sn seed particles in at least two general locations. Gray tin seed powder was prepared by temperature cycling of tin pieces with repeated transformation between the α-Sn and β-Sn phase, as was described previously.[17] In most cases, but not always, inoculation of tin drops with α-Sn seeds was done within a few hours after tin dripping.

## III. RESULTS AND DISCUSSION

### A. *Transformation of small tin drops on ML-coated sample*

Square Mo/Si-coated samples contaminated with one or several tin drops were tested first. The mirror samples were mounted on the sample holder, the tin pieces were loaded on the trip tray and the infection baskets were filled with α-Sn powder. The chamber was then pumped down to a vacuum pressure of $10^{-4}$ Pa or less. Samples were sometimes contaminated with one tin drop only. However, as described in the experimental section, a few pieces of β-Sn could be melted simultaneously on the tray by heating in parallel leading to sequential in-situ dripping of several tin drops. Fig. 5a shows a photo of a sample after tin dripping where several tin contamination spots were produced on its surface. One tin drop impacted the mirror at the lower right corner. A second drop hit the Al holder and produced several Sn fragments, with two large ones attaching to the upper part of the sample. A third tin drop ended up mainly on the holder outside of the view, but produced a series of small elongated tin strings across the middle section of the mirror sample. With these deposits this sample represented a case of fairly severe thick Sn contamination. According to image analysis, about 16.3 % of the mirror surface was covered by tin, leading to a presumed corresponding relative loss in EUV reflectivity of similar magnitude. To reach good contrast for imaging, the illumination with visible light was adjusted in such a way that the uncontaminated mirror surface produced a bright reflection. Only, the lower left corner of the substrate could not be illuminated in this way, creating a dark shadow on the mirror in this region.

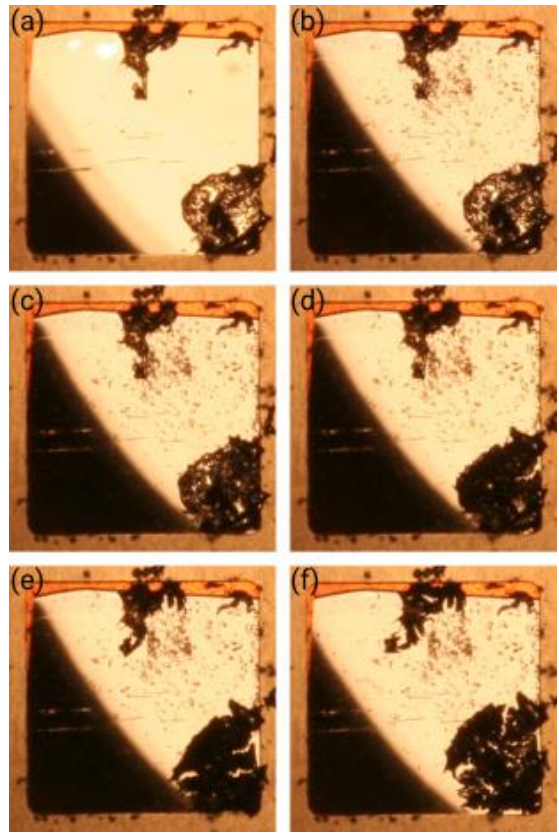

Fig. 5. (Color online) Images of ML-coated sample with small tin drop contaminations photographed at different stages during ongoing phase transformation: (a) after tin dripping, (b) after inoculation with seeds, after cooling for (c) 3 hours, (d) 4.5 hours, (e) 5.7 hours, and (f) 8 hours. The lower left part of the mirror was covered by a shadow.

The sample was infected with gray tin powder under uninterrupted vacuum conditions and at room temperature by positioning the basket by use of the manipulator rod above the EUV mirror and emptying it by rotation, thus dripping α-Sn seeds onto the surface near and on the largest contamination spots. Fig. 5b shows the sample after in-situ inoculation, with the additional gray tin powder being clearly visible in the image. After infection the sample holder was cooled down with cold nitrogen vapor to low temperatures such that a thermocouple wire clamped between the copper body of the sample table and the Al holder reached temperatures in the range of about -35 °C to -40 °C. Corresponding recordings at the exhaust of the cooling line showed temperatures down to between ~-50 °C and ~-60 °C. The reading at the holder was considered to be representative for the temperature of the sample. After about three hours of cooling the ongoing phase transformations of the tin pieces were clearly visible in the photographic images by subtle changes in light reflection and progressive darkening of the tin drops, see Fig. 5c. Deformation and volume expansion of the converted tin and corresponding small movements of Sn bulk material could then be noticed as well, leading to the development of a crack in the large tin drop after ~4 hours (Fig. 5d). More cracks developed and the transformed tin showed movements on the mirror surface caused by volume expansion, see Fig. 5e. As shown in Fig. 5f, after cooling for 8 hours, the larger tin drop contaminations on the mirror were essentially fully transformed to gray tin and exhibited numerous cracks where the mirror surface below became discernable. Sample cooling was then stopped.

On the next day the sample was removed from the chamber after slow venting with nitrogen gas. Fig 6a shows a photo of the contaminated mirror in the Al holder after removal of the sample holder unit from the chamber. Since the substrate had been at room temperature for more than 12 hours, the appearance of the tin drops had reverted back to a more whitish color, but the large drops were still very brittle. They were removed from the surface with ease by several gentle puffs of dry air. The sample appearance after tin cleaning is shown in the photo of Fig. 6b. The large tin drops could essentially be fully removed, since due to long enough cooling time the α-Sn phase had propagated through the entire bulk of contiguous Sn material. In Fig. 6b a small stripe of uncoated Si substrate is visible at the top of the sample. No differences were observed for coated and uncoated parts of the mirrors with respect to the behavior during tin contamination and transformation. Small pieces of tin debris adhering to the surface were more difficult to inoculate. A majority of the small tin strings on this sample had apparently not been infected by tin pest and still remained attached to the mirror. An image evaluation of the area coverage on the photo by Sn particles yielded a remaining coverage of 0.94 %. In contrast, all other test samples examined in our tests had smaller remaining tin coverages after cleaning, with corresponding fractions often below 0.1 %.

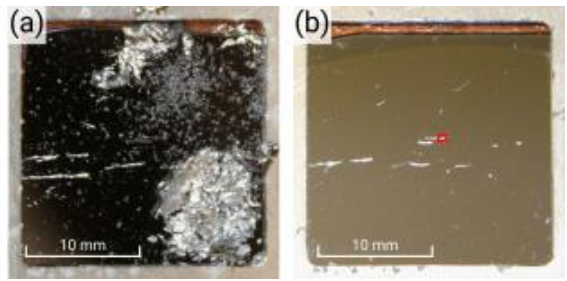

Fig. 6. (Color online) (a) Photo of mirror after in-situ transformation and removal from vacuum chamber. (b) Appearance of mirror after cleaning of tin pieces with dry air. The red rectangle near the center of the sample indicates the region examined in more detail in Fig. 7.

In agreement with the previous ex-situ experiments[17], cooling of the mirror samples to below -20 °C was found to be a prerequisite for reasonably fast progression of the phase transformation. A difference is that in the present in-situ transformation studies only the bottom side and the edges of the test substrates are cooled. The surface of the tin-drop contaminations and the seed particles where the transition is expected to start are also exposed to thermal radiation from the surrounding chamber walls. Therefore, their temperature is presumed to be slightly elevated compared to the substrate. In ex-situ transformation experiments inside of a freezer, on the other hand, the temperature is the same for the seeds on the surface, the tin drop and the substrate.

Apart from the temperature for cooling, a major additional factor for fast phase transition is the low amount of certain impurities present in the Sn material used here. Especially minute elemental impurities of Pb, Sb and Bi that are soluble in Sn can have a strong inhibiting influence on the phase transition. This was already pointed out in previous studies discussing the material science aspects of the transformation.[17,24] In the tests reported here we have used tin of 99.999 % purity where concentrations of inhibiting elements are expected to be in the range of only parts per million, or even less. In contrast to the prior ex-situ tests where tin was heated and dripped in air with associated strong surface oxidation of the droplet[17], the melting and dripping was now done under high-vacuum conditions where only a very thin oxidation layer can be expected to occur on the surface of the Sn drop. Since a surface oxidation layer can reduce epitaxial contact between seed and bulk material, it was previously often removed by application of HCl acid prior to infection with α-Sn. In the present study, this was not necessary for all conversion tests reported here (and also not possible as long as the samples were kept under vacuum).

A weakness of the method in its present developmental state is the relatively low inoculation efficiency at the start of the phase transformation. Many small seeds of α-Sn with different orientations and particle sizes were used to create a few nucleation sites of gray tin on a body of beta tin. Inspection of the seed particles used in this work with a light microscope revealed α-Sn particles with irregular shapes and sizes down to about 1 μm in diameter. In essence, given sufficient time for transformation, it seems to be more difficult presently to induce tin pest on many small separated Sn contamination pieces with small mass and size in comparison to large contiguous tin drops due to the nucleation hurdle for the initiation of the transformation.

For further analysis, the sample discussed above was inserted into a scanning electron microscope (SEM, Zeiss DSM 982 Gemini). Figure 7a shows one of the SEM images with low magnification (50x) for a tin string located in a region near the center of the sample, indicated by the red rectangle in Fig. 6b. The corresponding energy dispersive x-ray (EDX) spectrum displayed a strong and dominant Sn signal (Fig. 7b). A neighboring region visibly free from contamination in the SEM image gave an EDX intensity signal with only Si and Mo, originating from the Mo/Si-coating (see Fig. 7c), whereas the spectrum in Fig. 7d for a region half-covered by the tin string showed signals of Si, Mo and also Sn, as was to be expected. Some of the very small micrometer-size particles on the surface were analyzed by SEM-EDX with much higher magnification, as well. Apart from the predominant particles containing Sn, a few particles producing Al or Si signals could also be found. Small aluminum particles could have originated from the holder; Si dust might have come from the edges of the Si wafer substrate.

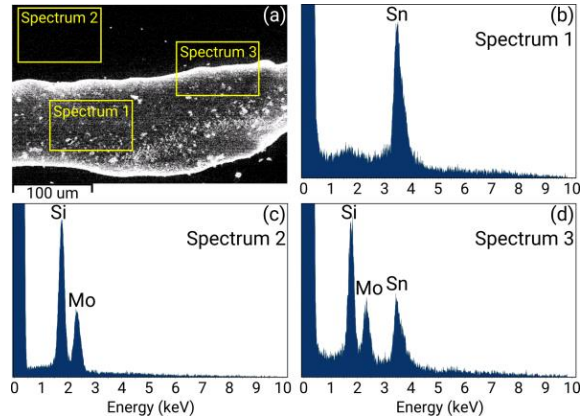

Fig. 7. (Color online) (a) SEM micrograph of red-labeled region of Fig. 6b. The regions where the corresponding EDX spectra 1 – 3 were recorded are indicated in the figure by yellow rectangles: (b) EDX spectrum 1, (c) EDX spectrum 2, (d) EDX spectrum 3.

## B.  *Transformation of massive tin drop on ML-coated sample*

Transformation experiments were also carried out with rectangular EUV mirror samples. Figure 8a shows such a sample where about 44% of the ML-coated mirror was covered by tin contaminations. Tin was dripped under vacuum onto this sample with two successive tin drops reaching the same spot. The first tin piece had only partially melted; the second drop covered the first one on the sample surface creating a massive Sn contamination, in part with a total height of almost 3 mm. In this case, the chamber was vented after tin dripping. The flange with the sample holder was then rotated by an angle of 54 degrees with respect to horizontal (see Fig. 8a), in order to create an inclination of the mirror. A large copper plate was placed as a tray at the bottom of the chamber below the sample holder to collect falling pieces of gray tin. The sample was inoculated by placing α-Sn powder on the top surface of its Sn contamination. Next, the chamber was pumped down again. After the vacuum pressure had reached levels of below $10^{-4}$ Pa, the sample holder was cooled with cold nitrogen vapor to temperatures of around -40 °C, as measured by a thermocouple probe attached under the sample holder next to the mirror. Photos of the sample were automatically recorded at regular intervals of ~5 minutes or longer. Fig. 8b shows the mirror at the start of in-situ cooling; Fig. 8c shows its appearance after more than 8 hours of cooling. Within a few hours most of the contamination was transformed, starting first in the infected area with accompanied strong volume expansion of the Sn material. Loose pieces of transformed gray tin fell successively off from the sample and down under their own weight onto the copper tray below, or they became trapped at the bottom side of the sample holder. After about close to 10 hours of cooling, the majority of the massive Sn contamination had already been transformed and detached from the top section of the mirror sample by falling down (see Fig. 8d). An exception was a transformed piece on the left side that was held in place by the sample holder. Magnification of the image revealed that the original Sn contamination was removed completely from the mirror surface (to within detection limit). Only the bare surface remained, showing some weak reflections of the surroundings. A string of tin in the lower section of the sample was also partially transformed to gray tin, with the transition starting from both of its ends and progressing further (see, for example, Fig. 8e).

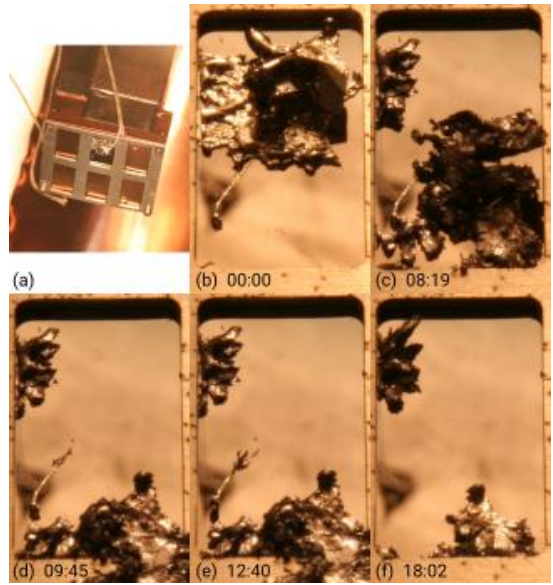

Fig. 8. (Color online) (a) Sample holder with ML-coated mirror contaminated by massive tin drop after tilting by an angle of 54°. (b – f) Images of sample with massive tin drop contamination photographed at different stages during cooling for phase transformation: (b) at start, (c) after ~8 hours, (d) after ~10 hours, (e) after ~13 hours, (f) after ~18 hours, corresponding to end of test. The total cooling time, counted from its start, is indicated.

At this point in time the cooling with cold nitrogen vapor was stopped. It was resumed after a pause of 12 hours on the next day. Within 2 hours of additional cooling, the remaining tin string was also almost fully transformed and fell off from the sample. No substantial changes of the tin contamination occurred after this point in time. Only converted gray tin was now left behind on the surface, prevented by the sample holder from falling down (see Fig. 8f). After venting of the chamber and removal of the Al sample holder, this loose tin contamination came off from the mirror immediately. Evaluation of the cleaned surface resulted in an area fraction coverage by remaining particles of only 0.06%. In agreement with the observation for all other tin-contaminated test samples examined, no coating damage was visible on this EUV mirror in analysis with a light microscope at moderate magnifications using 5x to 50x objectives. Due to strong volume expansion by structural changes during phase transformation, numerous cracks developed in the converted material and the adhesion of the imbrittled Sn contamination to the surface was greatly weakened. Deformation of the transformed tin also occurred, leading to separation of the α-Sn from the substrate surface at the interface. Therefore, the transformed brittle material fell off from the mirror under its own weight with ease, leaving a clean surface behind.

## C.  EUV reflectance after cleaning of ML-coated sample

The proof of successful cleaning can be obtained from EUV reflectance measurements of tin-cleaned samples. Tin is a strong absorber of EUV radiation. A homogeneous Sn layer on a EUV mirror with a thickness of only 5 nm would already lead to a (double-pass) transmission loss of about 50% for radiation at a wavelength of 13.5 nm.[25] For each contiguous microscopic Sn contamination with a thickness of more than ~0.1 μm a practically complete absorption of EUV radiation occurs in the covered area on the mirror. Furthermore, the expected normal-incidence reflectance of a thick flat layer of tin is negligibly small (only ~0.23 % at 13.5 nm)[25]. Thus, for the mirror samples studied here, all regions covered by Sn deposits are expected to give nearly zero contribution to the reflected EUV light intensity.

Potential EUV reflectance degradation of the optics was examined by using a pre-characterized mirror sample (25 mm x 25 mm square) that had been ML-coated with 50 Mo/Si- bilayers and a thin $SiO_2$ cap layer by magnetron sputter-deposition. The peak reflectance of the sample at near normal incidence that had been measured more than two years earlier at three positions on the mirror at the reflectometer of the PTB in Berlin[23], was $R_{max}$ = 68.0 % at a wavelength of 13.6 nm. A tin drop with a diameter of ~7.5 mm was dripped ex-situ onto this mirror. Even though oxidation of tin occurred during dripping, no use of HCl was made for removal of surface oxides. Less than 30 minutes later, after infection with α-Sn seed particles placed on the surface of the drop, the sample was moved to a freezer for transformation at -24 °C, as described previously[17]. Photos of the sample with the tin drop before cooling and after 7 hours of cooling are shown in Fig. 9. Essentially full phase transformation throughout the entire drop was reached within only about 7 hours of cooling. After cleaning of brittle gray tin fragments by puffs of dry air, the mirror surface was found to be free from any apparent Sn residues in visual inspection.

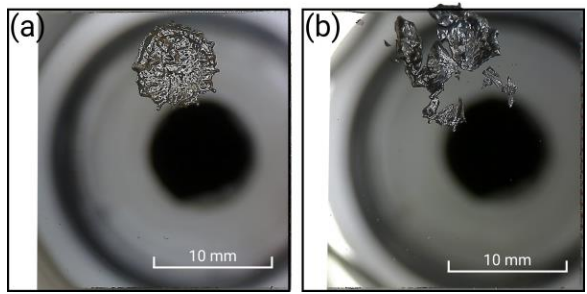

Fig. 9. Photo of ML mirror (a) after tin drop contamination and (b) after transformation at 7 hours of cooling. (The camera objective is reflected by the mirror.)

The reflectance of the cleaned sample was measured once more at PTB. As was briefly reported already[17], the peak reflectance in the region of the tin drop location was then found to be $R_{max}$ = 67.0 % at 13.6 nm. For a more detailed discussion, the corresponding EUV reflectivity data obtained at 5° incidence angle are compared in curves (1) and (2) of Figure 10a for the center of the mirror before and after the cleaning experiment, respectively. Additional reflectivity curves (3) and (4) shown in Fig. 10b (before sample use and after tin cleaning, respectively) were taken in regions of the sample that had been covered by the tin drop before its removal. An upper limit of 1% can be deduced for the EUV reflectance reduction of the mirror after tin dripping and cleaning. The decrease was more likely in the range of 0.5 % or less, since EUV reflectivity measurements in two regions outside of the tin contamination spot now resulted in peak reflectance values $R_{max}$ of 67.2 % and 67.7 %, respectively, also lower compared to the initial peak reflectance of 68.0 %. This may be attributed to small changes caused by aging of the coating and by degradation during sample handling. A very minor wavelength shift of the reflectivity curves by ~10 pm can be observed, also in the region that was not covered by the tin drop, when comparing the curves before and after sample cleaning (see Fig. 10). The full width at half maximum did not change significantly in comparison to the measurements before tin dripping.

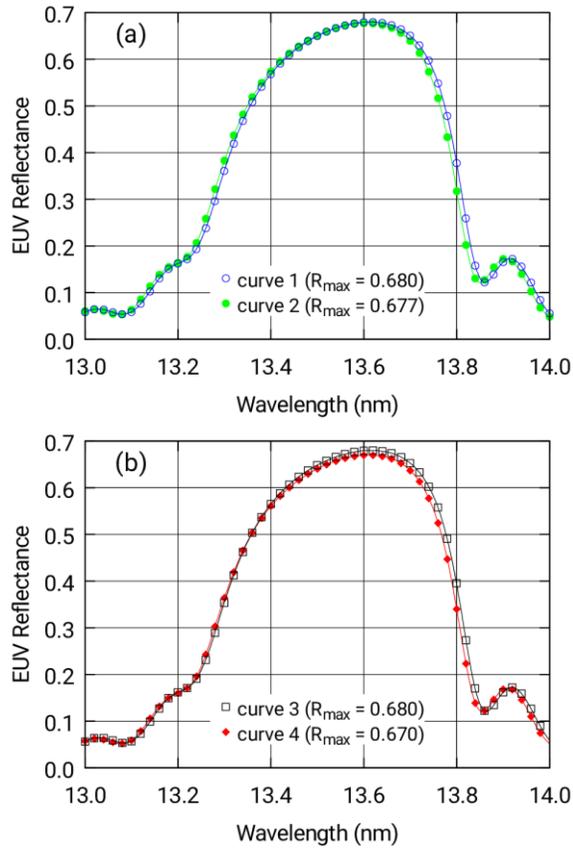

Fig. 10. (Color online) Wavelength dependence of EUV reflectance: Reflectivity curves before and after tin-cleaning of Mo/Si-coated mirror obtained (a) in the center of the sample (curves 1 and 2) and (b) in the region of tin drop (curves 3 and 4). Corresponding values of maximum reflectance for each curve are listed.

The result of EUV reflectivity mapping for a central portion of the sample at 5° incidence angle, taken by PTB at a fixed wavelength of 13.5 nm, i.e., 0.1 nm smaller than the wavelength of peak reflectance, is displayed in Fig. 11. The data show that the sample did not exhibit a fully homogeneous reflectance pattern. The gradient in reflectance for this scan at 13.5 nm was about 0.5 % across the inner portion of the cleaned sample, consistent with the observed variation of the peak reflectance values in the spots where reflectivity curves were recorded. The area that had been covered by the tin drop is indicated in the map, as well as the locations where reflectivity curves were measured. In the region covered by the tin drop the level of reflectance at 13.50 nm after tin cleaning of about 64.9% is of the same magnitude as the value obtained before cleaning at this wavelength (see Fig. 10 b). This coincidence can be explained by the observed very small wavelength shift of the curves after tin cleaning to slightly shorter wavelengths which effectively cancels the concurrent small decrease of reflectance. At the wavelength of 13.5 nm this led to almost the same EUV reflectance values before and after cleaning even though the peak reflectance of the reflectivity curves near 13.6 nm decreased slightly. It may be concluded that it is likely that a reduction in reflectance due to small remaining Sn particles on the mirror is considerably lower than the upper limit of 1%. It also has to be kept in mind that no clean-room environment was available for sample handling; therefore, very small amounts of dust accumulation, which could also cause a decrease in EUV reflectance, cannot be fully excluded.

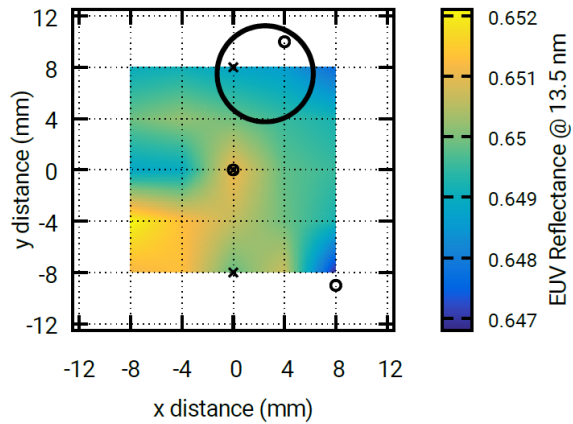

Fig. 11 (Color online) EUV reflectivity map after tin-cleaning recorded across the central part of the mirror surface at a fixed wavelength of 13.50 nm. The area that had been covered by the tin drop is indicated by a circle in the map. The locations on the mirror where reflectivity curves were measured before and after cleaning are also marked in the figure by crosses and open circles, respectively.

Apart from EUV absorption and scattering by Sn particles remaining on the mirror surface, other reflectance degradation mechanisms such as deterioration of the ML coating may be considered here for the sake of completeness. It is known that prolonged exposure to elevated temperatures during annealing can induce a reduction of EUV reflectance of Mo/Si ML mirror structures.[26] This occurs due to increased inter-diffusion and/or growth of interlayers at the ML interfaces, leading effectively to a reduction in contrast by increased layer transition zones. Such processes generally also cause small shifts in center wavelength of the ML reflectivity curves. The latter was clearly not observed here. Furthermore, it has to be kept in mind that during dripping of molten tin with temperatures of ~260 °C onto the ML samples even in a vacuum environment, the surface is affected by high temperatures only for a very short time, since the heat is dissipated in the substrate very quickly. Thus, the time is just not long enough for any significant inter-diffusion processes to take place in the coating. For comparison, for dripping onto a steel plate at room temperature, experiments and simulations for liquid tin drops of similar size in comparison to those produced here have shown that the plate surface reached its maximum temperature value of ~170 °C in just a few milliseconds.[27] The time for the Si substrate surface temperature to return to near equilibrium after the heat transfer from the Sn drop can be estimated to have a duration shorter than about 0.5 s due to the high thermal conductance of the ML-coated substrate. Considering potential effects of temperature changes from room temperature in the opposite direction during sample cooling, there is no deterioration mechanism of ML coatings known to us that could cause a reduction in EUV reflectance after extended cooling times of the mirrors to cryogenic temperatures. To simulate a severe test, we have immersed Mo/Si-coated Si-wafer samples in liquid nitrogen at a temperature of -196 °C for several minutes and also for more than 100 hours. At least no coating delamination or any other obvious visible damage was noticed during examination of the mirrors with a microscope after they were warmed up again back to room temperature.

## IV. SUMMARY AND CONCLUSIONS

The cleaning tests performed in this study on small Sn-contaminated EUV mirror samples have demonstrated fast and efficient in-situ cleaning of thick tin drops from the ML-coated substrates by induction of tin pest. It was found that after inoculation with α-Sn seed particles massive tin debris attached to mirror samples could be transformed into brittle gray tin. This change occurred generally within less than ~12 hours of substrate cooling with cold nitrogen vapor to temperatures of about -25 °C to -40 °C under high-vacuum conditions. Subsequently, tin contamination could easily be removed with dry or inert gas without any substantial Sn redeposition, requiring at most one venting cycle. For an inclined sample, it was shown that the converted gray tin fell off from the surface under its own weight. Our tests demonstrated that in-situ transformation and cleaning of Sn contaminated mirrors in an evacuated chamber led to results similar to those that were observed with ex-situ methods previously.[17] In-situ tin transformation and subsequent cleaning could be achieved within a day. Generally, the area fraction coverage of not transformed tin remaining on the sample surface after cleaning was found to be well below 1%, often even below 0.1%. Measurements of EUV reflectance curves for a pre-measured EUV mirror sample indicated a reflectivity decrease after tin dripping, phase transformation and cleaning in the range of only about 0.5%, with an upper limit of 1%.

We have tested a simple, low-cost cleaning method that applies specifically to tin contaminations, works fast for quite severe contamination by thick Sn deposits and introduces only very small amounts of high-purity gray tin powder into the vacuum environment. Having now established the capability for in-situ transformation and cleaning, this novel technique compares favorably to the other cleaning schemes mentioned in the introduction section. The use of acids is not required. During hydrogen plasma cleaning[13] that is dependent on the ion energy flux[28], there is a danger that the ML-coating of the mirror can be affected or modified by etching or by implantation of hydrogen. This cannot happen during cleaning by tin pest induction. The removal of very brittle Sn debris after phase transformation to α-Sn can be carried out at a fast rate and can likely be performed in a considerably more gentle way in comparison to expected behavior during $CO_2$ snow-jet cleaning of untransformed beta-tin deposits.

There are three main prerequisites required for application of this technique: Cooling to negative Celsius temperatures to below at least ~-20 °C is needed for macroscopic conversion of contamination of ductile β-Sn into brittle gray tin pieces at a reasonable speed. Some high-purity α-Sn seed particles are being introduced into the chamber. For the Sn contamination to be cleaned, the presence of very low impurity levels of potentially inhibiting elements that are soluble in tin is required to ensure fast phase transformation.

The implementation of this in-situ tin cleaning technique based on transformation by tin pest to contaminated collector mirrors in source modules of EUV scanners should be relatively straight-forward. The method can be applied especially to mirrors with internal substrate cooling channels where a mixture of ethylene glycol and water could be used as a low-temperature cooling fluid. The cleaning by allotropic phase transformation offers an interesting new way to reduce collector module maintenance times significantly. The cleaning technique by tin pest induction could also be applied to other optical elements, surfaces or internal hardware of source chambers for EUV lithography applications or to other tin-fueled light sources as long as the parts to be cleaned can be cooled to cryogenic temperatures for extended times.

As a next development step, the scalability of this cleaning technique to larger mirror areas and to mirrors with surface structure should be investigated. Other topics for future studies may be an examination of a possible thickness dependence in the cleaning of thin layers of deposited Sn and the influence on transformation speed by certain impurities that may be present in tin debris generated in EUV source chambers. A further important aspect is an improvement of the inoculation procedure to ensure that also contaminating tin particles of very small size and Sn micro-droplets on optical surfaces are efficiently converted to easily cleanable gray tin.

## ACKNOWLEDGMENTS

This study was motivated by the ongoing industrial development of EUV light sources. We are grateful to the molecular and surface physics group at Bielefeld University for general support and for supplying Mo/Si-coated EUV mirror samples. Furthermore, we would like to thank Torsten Feigl and his team at optiXfab for generously providing a pre-characterized ML-coated mirror sample and the group of Frank Scholze at PTB Berlin for carrying out the corresponding detailed EUV reflectance measurements. The help of Wiebke Hachmann in the SEM-EDX analysis of the samples at Bielefeld University is also gratefully acknowledged.

## References:

1. M. van de Kerkhof *et al.*, Proc. SPIE **10143**, 101430D (2017).
2. N. R. Farrar, B. M. La Fontaine, I. V. Fomenkov, D. C. Brandt, Adv. Opt. Tech. **1**, 279 (2012).
3. I. Fomenkov *et al.,* Adv. Opt. Tech. **6**, 173 (2017).
4. N. Farrar, D. Brandt, N. Böwering, Laser Focus World **45** (2009).
5. T. Feigl *et al.*, Proc. SPIE **8322**, 832217 (2012).
6. D. C. Brandt *et al.*, Proc. SPIE **9048**, 90480C (2014).
7. S. De Dea, M. Varga, A. I. Ershov, R. L. Morse, U.S. patent No. 9,073,098 (7 July 2015).
8. O. T. Ehrler, U. Meier, A. Uhl, H. Kierey, PCT application WO 2013/159928 A1 (31 October 2013).
9. T. Feigl *et al.*, Proc. SPIE **8679**, 86790C (2013).
10. M. M. J. W. van Herpen, D. J. W. Klunder, W. A. Soer, R. Moors, V. Banine, Chem. Phys. Lett. **484**, 197 (2010).
11. W. A. Soer, M. M. J. W. van Herpen, M. J. J. Jak, P. Gawlitza, S. Braun, N. N. Salashchenko, N. I. Chkhalo, V. Y. Banine, J. Micro-Nanolith. MEM **11**, 021118 (2012).
12. J. R. Sporre, D. Elg, D.N. Ruzic, S. N. Srivastava, I. V. Fomenkov, D. C. Brandt, Proc. SPIE **8679**, 86792H (2013).
13. D. Elg, J. R. Sporre, G. A. Panici, S. N. Srivastava, D. N. Ruzic, J. Vac. Sci. Technol. A **34,** 021305 (2016).
14. M. Moriya, Y. Ueno, T. Abe, A. Sumitani, U.S. patent application 2010/0025231 A1 (4 February 2010).
15. I. Varghese, C. W. Bowers, M. Balooch, Proc. SPIE **8166**, 816615 (2011).
16. M. Becker, U. Müller, O. Arp, PCT application WO 2015/043833 A1 (2 April 2015).
17. N. Böwering, Mater. Chem. Phys. **198**, 236 (2017).
18. G. A. Busch, A. Kern, Solid State Phys. **11**, 1 (1960).
19. E. Cohen, A. K. W. A. van Lieshout, Z. Phys. Chem. A **173,** 32 (1935).
20. W. J. Plumbridge, Circuit World **33**, 9 (2007).

21. A. D. Styrkas, Inorg. Mater. **39**, 806 (2003).

22. U. Kleineberg, T. Westerwalbesloh, W. Hachmann, U. Heinzmann, J. Tümmler, F. Scholze, G. Ulm, S. Muellender, Thin Solid Films **433**, 230 (2003).

23. C. Laubis *et al.*, Proc. SPIE **8679**, 867921 (2013).

24. M. Leodolter-Dworak, I. Steffan, W. J. Plumbridge, H. Ipser, J. Electron. Mater. **39**, 105 (2010).

25. B. L. Henke, E. M. Gullikson, L. C. Davis, Atom. Data Nucl. Data **54**, 181 (1993).

26. R. S. Rosen, M. A. Viliardos, M. E. Kassner, D. G. Stearns, S. P. Vernon, Proc. SPIE **1547**, 212 (1991).

27. M. Pasandideh-Fard, R. Bhola, S. Chandra, J. Mostaghimi, Int. J. Heat Mass Tran. **41**, 2929 (1998).

28. D. T. Elg, G. A. Panici, S. Liu, G. Girolami, S. N. Srivastava, D. N. Ruzic, Plasma Chem. Plasma P. **38**, 223 (2018).